\documentstyle[eqsecnum,aps]{revtex}

\begin{document}
\input epsf
\draft
\title{SPIN EFFECTS IN GRAVITATIONAL RADIATION
BACKREACTION\hskip5cm II. FINITE MASS EFFECTS}
 
\author{L\'aszl\'o \'A. Gergely, Zolt\'an I. Perj\'es and M\'aty\'as Vas\'uth}
\address
{KFKI Research Institute for Particle and Nuclear
Physics, Budapest 114, P.O.Box 49, H-1525 Hungary}
\maketitle
\begin{abstract}
 
A convenient formalism for averaging the losses produced by
gravitational radiation backreaction over one orbital period was
developed in an earlier paper. In the present paper we
generalize this formalism to include the case of a closed system
composed from two bodies of comparable masses, one of them having
the spin ${\bf S}$.
 
We employ the equations of motion given by {\em Barker} and
{\em O'Connell}, where terms up to linear order in the spin
(the spin-orbit interaction terms) are kept. To obtain the
radiative losses up to terms linear in the spin, the equations of
motion are taken to the same order. Then the magnitude $L$ of
the angular momentum ${\bf L}$, the angle $\kappa$ subtended by
${\bf S}$ and ${\bf L}$ and the energy $E$ are conserved.
The analysis of the radial motion leads to a new parametrization
of the orbit.
 
From the instantaneous gravitational radiation losses computed by
{\em Kidder} the leading terms and the spin-orbit terms are taken.
Following {\em Apostolatos, Cutler, Sussman} and {\em Thorne} (ACST),
the evolution of the vectors ${\bf S}$ and ${\bf L}$ in the
momentary plane spanned by these vectors is separated from the
evolution of the plane in space.
The radiation-induced change in the spin is smaller than
the leading-order spin terms in the momentary angular momentum
loss. This enables us to compute the averaged losses in the
constants of motion $E$, $L$ and $L_S=L\cos\kappa$. In the latter, the
radiative spin loss terms average to zero.
An alternative description using the orbital elements $a,e$ and $\kappa$
is given.
 
The finite mass effects contribute terms, comparable in magnitude,
to the basic, test-particle spin terms in the averaged losses.
 
\end{abstract}
 
\section{Introduction}
 
Coalescing binaries are important and copious sources of
gravitational waves
for the projected interferometric gravitational wave detection
experiments LIGO, VIRGO and LISA. The requisite signal templates
challenge theoreticians to predict the behaviour of such systems under
gravitational radiation backreaction. Much progress has been made in
this direction mainly by perturbative approaches. In the initial epoch,
the separation $r$ of the two bodies forming the binary is large
relative to the Schwarzschild radius and the relative motion can
be considered slow. This enables one to employ the post-Newtonian
expansion parameter $\epsilon\approx v^2\approx {m/r}$
\cite{KWW,Kidder}. Alternatively, with the system of units
chosen appropriately,
the inverse ${1/ c}$ of the speed of light can be used as an
expansion parameter\cite{Blanchet,BD}. Computations of
instantaneous radiation losses of energy, momentum and angular
momentum up to (post$)^{5/2}$-Newtonian order, including the
spin-orbit and spin-spin interaction terms, were given by
Kidder\cite{Kidder}.
 
The equations of motion of spinning bodies, disregarding radiation
backreaction, were
first considered by Barker and O'Connell\cite{BOC}, then by
Thorne and Hartle\cite{TH}. The relevant equations can be derived
from a generalized Lagrangian depending on relative position,
velocity and acceleration. The two-body problem is reduced to a
one-body problem by eliminating the center of mass. The
masses $m_1$ and $m_2$ still appear in the Lagrangian as reminders
of the initial two-body character of the problem. Apostolatos,
Cutler, Sussman and Thorne\cite{ACST} describe the evolution of
the spin and orbital angular momentum vectors in the presence of
radiation backreaction. They supplement the equations of motion
with the leading-order averaged gravitational radiation loss
terms, computed by Peters and Mathews\cite{PM,P}.
 
In the test particle limit the leading spin terms from the averaged
losses in energy, magnitude of the orbital angular momentum and spin
projection of the orbital angular momentum, respectively has been
computed by Ryan\cite{Ry2} for generic orbits.
Generalizing his earlier work on circular orbits\cite{Ry} and
Shibata's work on equatorial orbits\cite{Sh}, Ryan obtains the
first-order spin corrections to the Peters' equations of
radiation-induced change in the orbit parameters.
In a previous paper\cite{GPV}, to be referred to as {\bf I}, we
completed the description of the
test particle by giving the averaged losses in terms of unambigous
conserved quantities and by computing the radiation induced change
of the remaining orbit elements.
Our description relies on the use of Eulerian angle
variables, a new parametrization of the orbit and averaging by the
residue theorem.

 The computation of finite-mass effects can be carried out by choosing a
suitable spin supplementary condition ({\em SSC}). Three convenient
choices of {\em SSC} are discussed by Kidder\cite{Kidder}. A treatment
of the nonradiative evolution using a noncovariant {\em SSC} (A2b) of
\cite{Kidder}
can be found in Wex's work\cite{Wex}, and some radiative losses in
Rieth and Sch\"afer\cite{Rieth}.
 
The finite mass effects in the spin terms of averaged losses, using the
covariant {\em SSC},
were considered previously by Kidder, Will and Wiseman\cite{KWW} and by
Kidder\cite{Kidder}. They computed the energy loss and the
angular momentum loss for the case of circular orbits.
It is the purpose of the present
paper to obtain, for generic orbits, the averaged losses
of energy, magnitude of the orbital angular momentum and spin
projection of the orbital angular momentum.
 
We employ the second order Lagrangian suitable for the description
of the motion of comparable mass bodies in the presence of
spin, following Ref. \cite{BOC}. We want to
describe finite-mass effects in binaries consisting of a compact
body captured by another, massive, spinning body. The latter is
exemplified by a spinning black hole or neutron star. In this way
we generalize the picture of the evolution of a system consisting
of a black hole and a test particle, which we considered in
{\bf I}. The equations of
motion and the spin precession equations valid to the
(post$)^2$-Newtonian order are reviewed in the second section.
The constants of motion are the energy $E$ and the total angular
momentum ${\bf J}$. Since our goal is to get the radiation losses
up to the leading terms in the spin, we approximately picture the
spin precession by a rigid paralelogram
with sides $L$ and $S$, rotating about the total angular momentum
vector ${\bf J}$. In this approach the magnitude $L$ and
the spin projection $L_S$ of the orbital angular momentum are
constants of the motion. From their expressions an uncoupled radial
equation can be found.
 
In the third section, we determine the turning points. We then introduce
two  parameters, the 'eccentric
anomaly' and the 'true anomaly'. They have the respective
properties, and in the no-spin limit reduce to, the eccentric
anomaly and true anomaly of Kepler orbits.
The parametrization of the orbit by the eccentric anomaly enables
one to compute the orbital period defined as twice the period of time
elapsed between consecutive turning points.
 
In the fourth section, we compute the instantaneous radiation
losses. For this purpose, we make the assumption widely used in the
literature that the
radius $R$ of the spinning body is comparable with its mass $m_1$.
Then $S/r^2\approx m_1RV/r^2\approx \epsilon^2 V$ and
$L/r^2\approx \mu v/r\approx \epsilon^{3/2}$. Thus the spin
is of $\epsilon^{1/2}$ order smaller than the orbital angular
momentum. Under the assumption that the rotational velocity
$V$ is large, Apostolatos, Cutler, Sussman and Thorne\cite{ACST}
estimated the '$S$-changing piece of the radiation-reaction
torque' (the radiation loss of the spin) to be of order
$r\epsilon^{13/2}$. This is smaller by two full post-Newtonian
orders than the leading term in\cite{c1} $d{\bf J}/dt$ which is of order
$r\epsilon^{9/2}$, and by one half order than the spin-orbit
terms given by Kidder\cite{Kidder} which are of order
$r\epsilon^6V$. At the end of the section we present an estimate of the
order of magnitude of the radiation loss of the spin for smaller values
of $V$. We then find that, to the order
we are interested in, we may interpret the instantaneous total
angular momentum loss as consisting entirely of the loss in
the orbital angular momentum. These considerations enable us to
compute the loss of magnitude and spin projection of the orbital
angular momentum. However the latter will still contain
radiative spin loss terms, which are evaluated from the
Burke-Thorne potential. Fortunately, introducing suitable new
angular variables, all losses can be expressed in terms of the true
anomaly parameter alone.
 
In the fifth section we average these expressions by using
the residue theorem. We find that the spin-loss terms give no
averaged contribution to the losses. The finite-mass
backreaction effects are of the same order as the other spin
effects. The results obtained in the Lense-Thirring case emerge as
smooth limits, although the test particle case is outside the
framework of this paper.
 
The losses are given both in terms of conserved quantities and in
terms of 'ellipse parameters' $a,e$ and the angle $\kappa$ between
the spin vector ${\bf S}$ and the orbital angular momentum ${\bf
L}$. We give also the changes of the geometric orbit elements
$a,e$ and $\kappa$ due to radiation backreaction.
This is carried out in the sixth section.
 
As in ACST, we separate the motion of the spin
and orbital angular momentum vectors in their momentary planes
from the motion of the plane. Our approximation in describing
the radiation effects on the former motion up to linear terms in
the spin.
 
Since we eliminate the post-Newtonian parameter $v^2$ from
the losses at the very beginning of the averaging process,
we keep the gravitational constant $G$ and the speed of light $c$
in our formulae. Our post-Newtonian parameter is
$\epsilon\approx v^2/c^2\approx Gm/c^2r$. The explicit $c$
dependence allows an easier bookkeeping in the post-Newtonian
expansions.

\section{The Orbit of the Binary System}
 
We consider the motion of a bound two-body system with masses $m_1$
and $m_2$ and spins $\bf S$ and $0$, respectively.
Considering only the leading-order spin-orbit coupling, and adopting the
spin supplementary condition of\cite{KWW}, the Lagrangian is:
\begin{equation}
{\cal L}=\frac{\mu{\bf v}^2}{2}+{G m \mu\over r}+\delta{\cal L}\ ,
\end{equation}
where
$r=\vert{\bf r}\vert$ is the relative distance,
${\bf v}$ is the relative velocity,
and the perturbation term due to rotation effects is
\begin{equation}
\label{eq:delag}
\delta {\cal L}={2(1+\eta)G\mu\over c^2r^3}{\bf v}({\bf
r}\times{\bf S})
               +{\eta\mu\over 2c^2m}{\bf v}({\bf a}\times{\bf S})
\ .\label{delag}\end{equation}
Here $\mu$ is the reduced mass and $m$ the total mass of the system,
\begin{equation}
\mu={m_1 m_2\over m_1+m_2}\ ,\qquad m=m_1+m_2\ ,\qquad
\eta={m_2\over m_1} \ .
\end{equation}
The Lagrangian depends on the
relative acceleration ${\bf a}$. The application of a generalized,
second-order variational formalism yields the equations of motion:
\begin{equation}
\label{eq:acc}
{\bf a}=-{Gm\over r^3}{\bf r}+{G\over c^2r^3}
\left\{{6(1+\eta)\over r^2}
           {\bf r}\left[({\bf r}\times{\bf v})\cdot{\bf S}\right]
       -(4+3\eta){\bf v}\times {\bf S}
       +3(2+\eta){\dot r\over r}{\bf r}\times {\bf S} \right\}\ ,
\label{acc}
\end{equation}
and the momenta:
\begin{equation}
\label{eq:q}
{\bf q}={\partial {\cal L}\over \partial{\bf a}}
       ={\eta\mu\over 2c^2m}{\bf S}\times {\bf v} \ ,
\label{q}
\end{equation}
\begin{equation}
\label{eq:p}
{\bf p}={\partial {\cal L}\over \partial {\bf v}}-{\bf \dot q}
       =\mu {\bf v}+{(2+\eta)G\mu\over c^2r^3}{\bf r}\times {\bf S} \
. \label{p}
\end{equation}
The energy $E$ and the total angular momentum ${\bf J}$ are
constants of motion:
\begin{equation}
\label{eq:energy}
E={\bf p}\cdot {\bf v}+{\bf q}\cdot {\bf a}-{\cal L}
 =\frac{\mu v^2}{2}-{G m \mu\over r}+{\eta G\mu\over
c^2r^3} {\bf S}\cdot ({\bf r}\times{\bf v}) \ ,
\label{energy}\end{equation}
\begin{equation}
\label{eq:J}
{\bf J}={\bf S}+{\bf L} \ .
\label{J}
\end{equation}
Here the orbital angular momentum ${\bf L}$ is
\begin{equation}
\label{eq:Lvect}
{\bf L}={\bf r}\times {\bf p}+{\bf v}\times {\bf q}
       ={\bf L_N}+{\bf L_{SO}}
\label{Lvect}
\end{equation}
with the Newtonian orbital angular momentum
\begin{equation}
{\bf L_N}=\mu {\bf r}\times{\bf v}\
\end{equation}
and the spin-orbit term
\begin{equation}
{\bf L_{SO}}={\mu\over c^2m}
\left\{(2+\eta){Gm\over r^3}
              \left[{\bf r}\times ({\bf r}\times{\bf S})\right]
-{\eta\over 2}\left[{\bf v}\times ({\bf v}\times{\bf S})\right]
\right\}\ .
\end{equation}
Formally the spin terms arising in these expressions are of order
$\epsilon S$. Since the spin is of order $S\approx r^2\epsilon^2 V$,
the terms containing the spin in the Lagrangian, the
acceleration, the energy and the orbital angular momentum are of order
$\epsilon^{3/2}V/c$ higher than the corresponding Newtonian
terms.
 
The spin precession equation\cite{KWW} is:
\begin{equation}
\label{eq:Sprec}
{\bf \dot S}=(4+3\eta){G\over 2c^2r^3}{\bf L_N}\times{\bf S}\ .
\label{Sprec}
\end{equation}
Hence the magnitude $S$ of the spin is constant. Note that up to
this point the description is valid up to (post)${}^2$-Newtonian
order.
 
Since we are interested in the leading spin terms of the
gravitational radiation backreaction losses, we may replace
${\bf L_N\to J}$ by inserting terms of higher order in $S$.
This is allowed, as we are neglecting terms of order
$\epsilon S^2=\epsilon^5$ in the precessional angular velocity,
which are $\epsilon^{3/2}$ order higher then the spin terms that
we have kept. (We shall return to the order-of-magnitude estimates
at the end of this section.)
 
From the conservation of total angular momentum (\ref{J})
and the (\ref{Sprec}) precession of ${\bf S}$ about ${\bf J}$,
the vector ${\bf L}$ is found to precess about ${\bf J}$:
\begin{equation}
\label{eq:Lprec}
{\bf \dot L}=-{\bf \dot S}=(4+3\eta){G\over 2c^2r^3}
{\bf J}\times{\bf L}\
. \label{Lprec}\end{equation}
   Combining (\ref{eq:Sprec}) (with ${\bf J}$ instead of ${\bf
L_N}$) and (\ref{eq:Lprec}) it follows that the
angle $\kappa$ subtended by ${\bf S}$ and ${\bf L}$ is constant.
As shown in Appendix A, the angle $\kappa$ is constant to order
$\epsilon^{3/2}$.
Thus we have the picture of a paralelogram of constant
sides $L$ and $S$ rigidly rotating about its diagonal ${\bf J}$ (Fig.1).
 
     We choose the Cartesian coordinates ${\bf r}=\{x,y,z\}$ of
the reduced-mass particle with origin at the center of mass such that
${\bf J}$ points along the $z$ axis\cite{c2}
. In this approximation we can choose the conserved quantities
characterizing the orbital motion as: the energy $E$ (due to the
time-independent nature of the Lagrangian), the magnitude $L$ of
${\bf L}$ (the vector ${\bf L}$ undergoes a pure rotation in the order
we are considering) and $L_S$, the spin-oriented component of
${\bf L}$ ({\em cf.} Appendix A):
\begin{equation}
\dot E=\dot L_S =\dot L=0\ .
\end{equation}
This description is close to the one in
the Lense-Thirring approximation. The inclusion of finite mass
does not destroy the basic features of the formalism
developed in the test particle limit.
In computing the expressions of the constants of motion, ${\bf
L_N}\cdot{\bf S}$ was replaced by $L_SS$, which is compatible
with the order we are considering.  Using polar coordinates,
$x=r\sin\theta\cos\varphi$, $y=r\sin\theta\sin\varphi$,
$z=r\cos\theta$, the constants of the motion are:
\begin{eqnarray}
\label{eq:E}
E&=&{\mu v^2\over 2}-{G m \mu\over r} + \eta {GL_SS\over c^2r^3}\
=\
{\frac{\mu}{2}[\dot r^2+r^2(\dot\theta^2+\sin^2\theta\ \dot\varphi^2)]}
-{G m \mu\over r} + \eta {GL_SS\over c^2r^3} \ ,
\label{E}\\
\label{eq:L2}
L^2&=&{\mu^2}r^4(\dot\theta^2+\sin^2\theta\ \dot\varphi^2)
-4{G\mu L_SS\over c^2r} + {2\eta\over c^2m}EL_SS \ ,
\label{L2} \\
\label{eq:LS}
L_S&=&{\bf L}\cdot {{\bf S}\over S}\ =\ L\cos{\kappa} \ .
\label{LS}
\end{eqnarray}
From the first expression of the energy, $v^2$ can be expressed in
terms of constants of motion and $r$:
\begin{equation}
\label{eq:v2}
v^2={2\over\mu}E + {2G m\over r} - \eta {2GL_SS\over c^2\mu r^3}
\ .
\label{v2} \\
\end{equation}
From the second expression of the energy in (\ref{eq:E}) and from
(\ref{eq:L2}), the equation for the radial motion is found to
decouple from the angular degrees of freedom:
\begin{equation}
\label{eq:radial}
\dot r^2=2\frac{E}{\mu}+2\frac{Gm}{r}-\frac{L^2}{\mu^2r^2}
+2\eta {EL_SS\over c^2m\mu^2r^2}
-2(2+\eta)\frac{GL_SS}{c^2\mu r^3} \ .
\label{radial}
\end{equation}
We conclude this section by considering some order-of-magnitude
estimates in the approximation which has led to the picture of a
rigidly rotating paralelogram. The neglected time-dependent terms in
$L^2$ and $L_S$ are
$(4+3\eta)\int {\bf L}\cdot ({\bf L_{SO}}\times {\bf S})/r^3
 dt\approx r^4\epsilon^{6}$ and
$(4+3\eta)/2 \int {\bf L}\cdot ({\bf S}\times {\bf L_{SO}})/r^3S
 dt\approx r^2\epsilon^4$, respectively. These are of order
$\epsilon^{3/2}$ and $\epsilon$ smaller than the terms we keep.
In the radial equation (\ref{eq:radial}), as declared, we keep terms
linear in the spin and drop terms of order $\epsilon^{3/2}$ smaller.

\section{Parameterization of the orbit}
 
 In this section, we employ Eq. (\ref{eq:radial})
for determining the turning points and then finding convenient
parametrizations of the orbit. We describe the procedure
only in its outlines, as a detailed exposition was
previously given for the Lense-Thirring case in {\bf I}.
The condition for turning poits $\dot r=0$ yields a cubic
equation. By
looking for roots slightly different from the roots known for the
nonspinning case, one unphysical root and the turning points
$r_{{}^{max}_{min}}$ are found:
\begin{equation}
\label{eq:turning}
r_{{}^{max}_{min}}=
-{Gm\mu\over 2E} + {(2+\eta )G\mu L_SS\over c^2L^2}
        \mp\Bigl[{A_0\over 2E}
               + {(2+\eta )G^2m\mu^2 L_SS\over c^2L^2A_0}
               - \eta{EL_SS\over c^2m\mu A_0}\Bigr]\ .
\label{turning} \end{equation}
Here $A_0$ is the length of the Runge-Lenz vector to the zeroth
order in the spin:
\begin{equation}
\label{eq:A0}
A_0=\sqrt{G^2m^2\mu^2+{2EL^2\over\mu}}\ .
\label{a0}\end{equation}
As in the Lense-Thirring case, we define the eccentric anomaly
parametrization of the Kepler orbits,
\begin{equation}
\label{eq:xi}
r=-{Gm\mu\over 2E} + {(2+\eta )G\mu L_SS\over c^2L^2}
        +\Bigl[{A_0\over 2E}
               + {(2+\eta )G^2m\mu^2 L_SS\over c^2L^2A_0}
               - \eta{EL_SS\over c^2m\mu A_0}\Bigr]\cos\xi\ ,
\label{xi}\end{equation}
such that $r_{{}^{max}_{min}}$ is at $\xi=\pi$ and $\xi=0$,
respectively. The parameter derivative $dr/d\xi$ follows from
(\ref{xi}). Expressing $1/\dot r$ from (\ref{radial}) and
linearizing in $S$, the expression for $dt/d\xi$ is found:
\begin{equation}
\label{eq:txi}
{dt\over d\xi}={1\over\dot r}{dr\over d\xi}=
{\mu^2(Gm\mu-A_0\cos\xi)\over (-2\mu E)^{3\over 2}} +
{((2+\eta)G^2m^2\mu^3-\eta EL^2)L_SS\cos\xi
\over c^2mL^2A_0(-2\mu E)^{1\over 2}} \ .
\label{txi}\end{equation}
 
Integration of (\ref{txi}) from 0 to $2\pi$ gives the orbital
period:
\begin{equation}
\label{eq:period}
T=2\pi{Gm\mu^3\over (-2\mu E)^{3\over 2}} \ .
\label{period}\end{equation}
Note that the expression for the period has the same
functional form as in the Lense-Thirring case.
 
The eccentric anomaly parametrization $r=r(\xi)$ was useful
in computing the orbital period, but it leads to unnecessarily
complicated expressions when instantaneous losses need to be
averaged. For the latter purpose we introduce the true anomaly
parametrization $r=r(\chi)$ requiring that it has the following
properties:
\begin{eqnarray}
(a)&{}& \quad r(0)=r_{min}\quad and\quad r(\pi)=r_{max}\\
(b)&{}& \quad {dr\over d(\cos\chi)}=-(\gamma_0+S\gamma_1)r^2
\end{eqnarray}
where $\gamma_0,\gamma_1$ are constants. Property (b)
generalizes Kepler's second law for the area. The unique
parametrization satisfying both (a) and (b) is:
\begin{eqnarray}
\label{eq:chi}
r=\frac{L^2}{\mu (Gm\mu+A_0\cos\chi)}
&+&\frac{2(2+\eta)GL_SS}{c^2L^2A_0}\
\frac{A_0(2G^2m^2\mu^3+EL^2)+Gm\mu(2G^2m^2\mu^3+3EL^2)\cos\chi}
{(Gm\mu+A_0\cos\chi)^2} \nonumber\\
&-&\frac{2\eta EL_SS}{c^2m\mu^2A_0}\
\frac{Gm\mu^2A_0+(G^2m^2\mu^3+EL^2)\cos\chi}
{(Gm\mu+A_0\cos\chi)^2} \ .
\label{chi}
\end{eqnarray}
In the same way as $dt/d\xi$ [Eq. (\ref{txi})] was obtained,
$dt/d\chi$ is found to have the form
\begin{equation}
\label{eq:tchi}
{dt\over d\chi}={1\over\dot r}{dr\over d\chi}=
{\mu r^2\over L}\Bigl[1
-{L_SS\over c^2mL^4}
\Bigl((2+\eta)Gm\mu^2(3Gm\mu+A_0\cos\chi)-\eta E L^2\Bigr)\Bigr] \ .
\label{tchi}
\end{equation}
The true anomaly parametrization (\ref{chi}) will be used for
the integration over one period of all instantaneous losses.
In the next section, all of these losses will
be expressed in the form $F=F(\chi)$. They
contain no other $\chi$ dependent factor in the denominator
than $r^{2+n}$, where $n$ is positive integer.
Time integration can be replaced by parameter integration;
\begin{equation}
\int_0^{T}F(t)dt=
\int_0^{2\pi}F(\chi){dt\over d\chi}\ d\chi \ ,
\end{equation}
where $dt/d\chi$ is given by (\ref{eq:tchi}). Here we encounter
an advantage
of property (b) of our parametrization: the $\chi$
dependence of the denominator of the integrand is especially
simple, $r^n$.
 
As in the Lense-Thirring case, the integrals are evaluated
by computing the residues enclosed in the
circle $\zeta=e^{i\chi}$.
We find, as the second advantageous feature of the
parametrization that there is only one pole, at $\zeta=0$.
Averaging over one period is achieved
by dividing the result by the period $T$ [Eq.(\ref{period})].
 
\section{Instantaneous radiative losses}
 
 In this section, we obtain the instantaneous losses in the energy
$E$, magnitude
$L$ and spin projection $L_{S}$ of the orbital angular momentum.
We then find suitable angular variables such that all losses can be
rewritten in terms of the true anomaly parameter $\chi$.
 
From the radiative power and the instantaneous total angular momentum
losses of the binary system given by Kidder\cite[formulae
(3.24),(3.28)] {Kidder}, we keep the Newtonian and spin-orbit
terms: \begin{eqnarray} \label{eq:dEdt}
{dE\over dt}=&-&{8\over 15}{G^3m^2\mu^2\over c^5r^4}\,(12v^2-11\dot{r}^2)
 -{8\over 15}{G^3m\mu\over c^7r^6}\,({\bf L}_N\cdot{\bf S})
\Bigl[27\dot{r}^2-37v^2-12{Gm\over r}
     +\eta\Bigl(51\dot{r}^2-43v^2+4{Gm\over r}\Bigr)\Bigr] \\
  \label{eq:dJdt}
{d{\bf J}\over dt}=&-&{8\over 5}{G^2m\mu\over c^5r^3}
            {\bf L}_N\Bigl(-3\dot{r}^2+2v^2+2{Gm\over r}\Bigr)
  -{4\over 5}{G^2\mu^2\over c^7r^3}\Bigl\{
    -{2\over 3}{Gm\over r}(\dot{r}^2-v^2)(1-\eta){\bf S}
    -\dot{r}{Gm\over 3r^2} {\bf r}\times({\bf v}\times{\bf S})(7+5\eta)
    \nonumber\\
   &+& {Gm\over r^3}{\bf r}\times({\bf r}\times{\bf S})
        \Bigl[6\dot{r}^2-{17\over 3}v^2+2{Gm\over r}
            +\eta\Bigl(9\dot{r}^2-8v^2-{2\over 3}{Gm\over r}\Bigr)\Bigr]
\nonumber\\
   &+& {\dot{r}\over r}{\bf v}\times({\bf r}\times{\bf S})
        \Bigl[-30\dot{r}^2+24v^2+{29\over 3}{Gm\over r}
            +5\eta\Bigl(-5\dot{r}^2+4v^2+{5\over 3}{Gm\over r}\Bigr)\Bigr]\\
   &+& {\bf v}\times({\bf v}\times{\bf S})
        \Bigl[18\dot{r}^2-12v^2-{23\over 3}{Gm\over r}
            +\eta\Bigl(18\dot{r}^2-{35\over 3}v^2-9{Gm\over r}\Bigr)\Bigr] \nonumber\\
   &+& {{\bf L}_N\over \mu^2r^2}({\bf L}_N\cdot {\bf S})
        \Bigl[30\dot{r}^2-18v^2-{92\over 3}{Gm\over r}
            +\eta\Bigl(35\dot{r}^2-19v^2-{71\over 3}{Gm\over r}\Bigr)\Bigr] \Bigr\}
\nonumber
\end{eqnarray}
 We want to compute the
instantaneous radiative losses of the constants of motion.
As we argued in the Introduction and will show in detail later in
this section, the total angular
momentum loss can be taken equal to the orbital angular momentum
loss, $d{\bf J}/dt=d{\bf L}/dt$. The
loss in the magnitude of the orbital angular momentum is given by
\begin{equation}
2L{d L\over dt}={d L^2\over dt}={d(L_iL_i)\over dt}
                                =2L_i{d L_i\over dt}\ .
\end{equation}
The loss in $L_S$ is given as:
\begin{equation} \label{LSloss}
{d L_S\over dt}={d \over dt}\left({\bf L}\cdot{{\bf S}\over S}\right)
               ={d{\bf L}\over dt}\cdot{{\bf S}\over S}
                +{{\bf L}\over S} \cdot{d{\bf S}\over dt}
          -\left({{\bf L}\over S} \cdot{{\bf S}\over S}\right)
           \left({{\bf S}\over S}\cdot{d{\bf S}\over dt}\right)\ .
\end{equation}
Although the radiation-induced change in the spin ${\bf S}$ is
$\epsilon^{1/2}$ order smaller than the spin-orbit part of
the change in the orbital angular momentum, due to the vectors
${\bf L}/S\approx\epsilon^{-1/2}$, the second and third terms
are of comparable order with the first term.
We will evaluate the terms depending on the
structure of the spinning body later in this section.
The first term is found simply by multiplying (\ref{eq:dJdt})
by ${\bf S}/S$.
 
The next step is to find variables in which these
radiative losses can be expressed solely in terms of the parameter
$\chi$ and constants of the motion. The polar coordinates
$(\theta,\phi)$
are not suitable for this purpose: $\theta$ is not a monotonous
function of time, thus the root of $\dot\theta^2$ can not be
extracted unambiguously.
 
We introduce new variables conveniently characterizing
the separation vector ${\bf r}$, the spin vector ${\bf S}$ and the
orbital angular momentum vector ${\bf L}$.
The relative coordinates can be expressed in terms of the time
dependent Euler angles $\Psi,\iota_N,\Phi$ as:
\begin{eqnarray}\label{eq:xyz}
x&=&r(\cos\Phi\cos\Psi-\cos\iota_N\sin\Phi\sin\Psi)\ ,\nonumber\\
y&=&r(\sin\Phi\cos\Psi+\cos\iota_N\cos\Phi\sin\Psi)\ ,\\
z&=&r \sin\iota_N\sin\Psi \ .\nonumber
\end{eqnarray}
Here $\iota_N$ is the polar angle and $\Phi-\pi/2$ the
azimuthal angle of the Newtonian orbital momentum ${\bf L}_N$
(Fig. 1).
The variable $\Psi$ is the angle subtended by the momentary
position vector and the node line
(the intersection of the the plane orthogonal to the ${\bf L}_N$
with the plane orthogonal to the total angular momentum ${\bf J}$).
From the leading terms of the equations of motion (Cf.
Appendix of  {\bf I}) it follows that
$\Psi$ is a monotonous function of $t$. Thus we extract
the square root of the equation for $\dot\Psi^2$, choosing the
positive root.
 
\begin{figure}[htb]
\epsfysize=7cm
\centerline{\hfill
\epsfbox{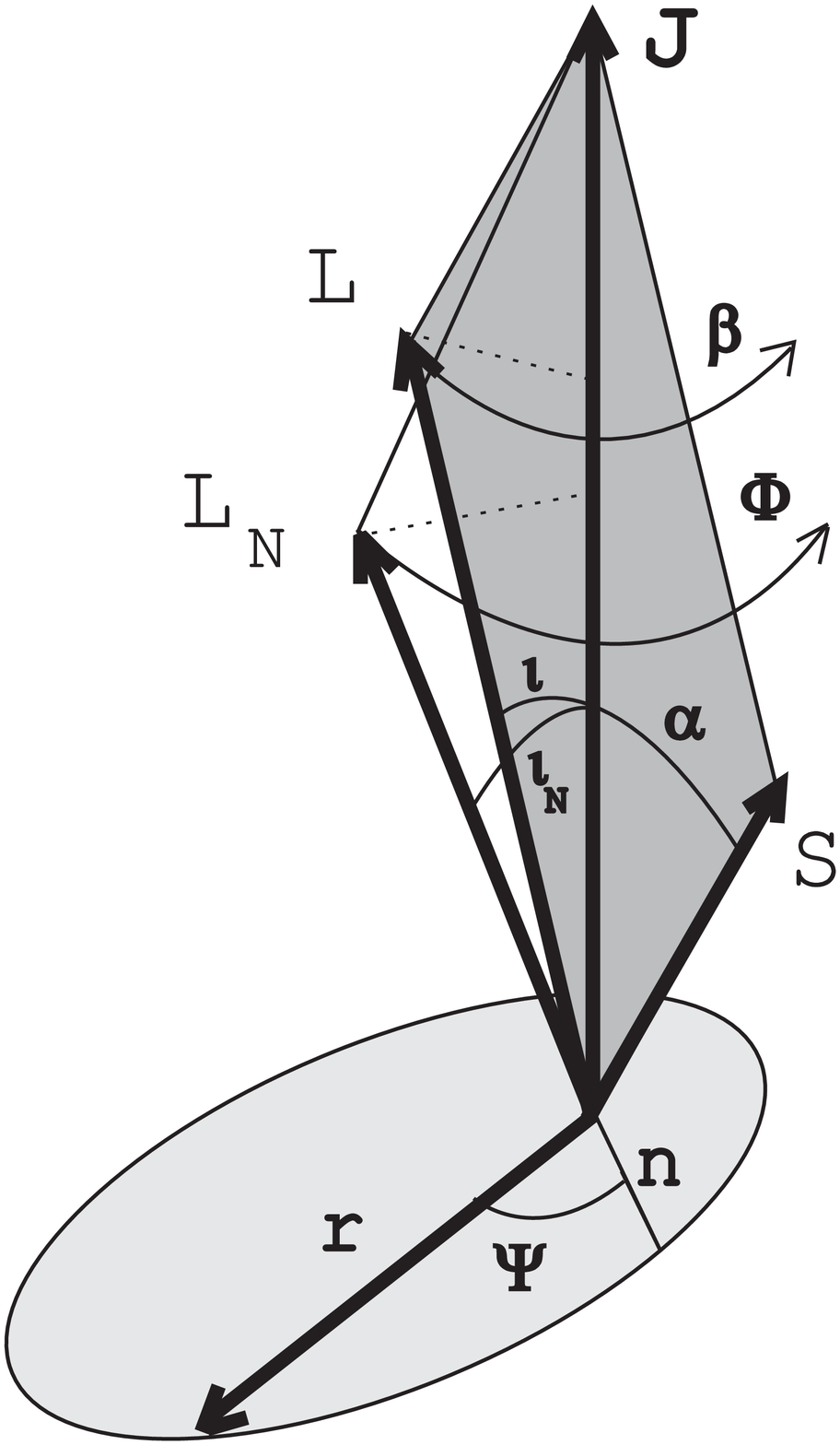}\hfill}
\caption{The orientation of the orbit and the angular momenta.}
\end{figure}
 
In the absence of spin, the orbital plane is at rest, and
$\Psi$ is the usual polar angle, satisfying the area law.
Thus the time derivatives of the angles are given by
\begin{equation}
\dot\Psi={\frac{L}{\mu r^2}}+S \dot\Psi_1\ ,\qquad
\dot\Phi=S \dot\Phi_1\ ,\qquad
\dot\iota_N=S \dot\iota_1
\ .\end{equation}
We will not need the explicit expressions of the first-order (in $S$)
terms $\dot\Psi_1, \dot\Phi_1$ and $\dot\iota_1$ for parameterizing
the instantaneous losses. Rather it will be sufficient for us to
find how they are interrelated. First, by computing the square $L_N^2$
of the Newtonian orbital angular momentum in two different ways:
\begin{equation}
\label{eq:Ln2}
L_N^2\ =\ (\mu {\bf r}\times {\bf v})^2
     \ =\ L^2-2{\bf L_N}\cdot{\bf L_{SO}}\ ,
\end{equation}
where $2{\bf L_N}\cdot{\bf L_{SO}}$ are the spin terms from
(\ref{L2}), the following relation is found:
\begin{equation}
\label{eq:rel1}
\dot\Psi_1=-\cos\iota_N\dot\Phi_1 + \cos\kappa{2Gm\mu-\eta Er\over
 c^2m\mu r^3}\ .
\end{equation}
 
Second, we compute the $z$ component of ${\bf L_N}$ in two
independent ways:
\begin{equation}\label{LNz}
(L_N)_z\ =\ (\mu {\bf r}\times {\bf v})_z
       \ =\ (L^2-2{\bf L_N}\cdot{\bf L_{SO}})^{1\over2}\cos\iota_N
\ . \end{equation}
Linearizing the second equality of (\ref{LNz}) in the spin, we obtain
\begin{equation}
\label{eq:rel2}
\dot\Phi_1={\tan\Psi\over\sin\iota_N}\dot\iota_1\ .
\end{equation}
When the relations (\ref{eq:rel1}) and (\ref{eq:rel2}) are
inserted in the instantaneous losses of energy, magnitude and
spin projection of the orbital angular momentum,
all terms with $\dot\Psi_1, \dot\Phi_1, \dot\iota_1$ will cancel.
 
The paralelogram spanned by the spin and orbital angular momentum
is next described in terms of the new variables (Fig.1). We denote the
polar and azimuthal angles of ${\bf S}$ and ${\bf L}$ by
$\alpha,\beta$ and $\iota,\beta+\pi$. During the first
period when ${\bf J}$ lies on the $z$ axis, $\kappa=\iota+\alpha$.
Similarly as with ${\bf L}$ and ${\bf L_N}$,
the azimuthal and polar angles differ only by first order terms
in the spin. As a result, no dependence on the azimuthal angles
$\Phi$ and $\beta$ remains in the losses.
 
As we mentioned in the Introduction, the spin is of $\epsilon^{1/2}$
order smaller than the orbital angular momentum. This implies that
the side $S$ of the paralelogram shrinks.
As the angles $\iota,\ \iota_N$ and $\alpha$ are present in the
losses only in the spin-terms, we need only their zeroth order
expressions:
\begin{equation}
\iota_N=\iota=0,\quad \alpha=\kappa\ .
\end{equation}
The nonradiative evolution of the angles $\iota$ and $\iota_N$
is discussed in Appendix B.
 
  From (\ref{v2}) and (\ref{radial}) $v^2$ and $\dot r^2$ can be
replaced by expressions containing solely $r$ and constants.
Making use of the foregoing observations, we obtain the
instantaneous losses of energy,
magnitude and spin component of the orbital angular momentum:
\begin{eqnarray} \label{instloss}
{dE\over dt} =
&-& {8\ G^3m^2\over 15c^5r^6}
\left(2\mu Er^2+2Gm\mu^2r+11L^2\right)                       \nonumber\\
&+& {8G^3mLS\cos\kappa\over 15c^7\mu r^8}
\Bigl[20\mu Er^2-12Gm\mu^2r+27L^2
      +\eta\left(6\mu Er^2-18Gm\mu^2r+51L^2\right)\Bigr]     \nonumber\\
{dL\over dt} =
&+& {8G^2mL\over 5c^5\mu r^5}\left(2\mu Er^2-3L^2\right)     \nonumber\\
&+& {8G^2S\cos\kappa\over 15c^7\mu^2r^7}
 \Bigl\{12Gm\mu^3Er^3+3\mu r^2\left(6EL^2
   + G^2m^2\mu^3\right)-11Gm\mu^2L^2 r                       \nonumber\\
& &\qquad\qquad\quad +\eta\left[2\mu^2E^2r^4+12Gm\mu^3Er^3+
     3\mu r^2(5EL^2 + G^2m^2\mu^3)-5Gm\mu^2L^2 r+15L^4\right]\Bigr\}  \\
{dL_S\over dt} =
&+&{8G^2mL\cos\kappa\over 5c^5\mu r^5}\left(2\mu Er^2-3L^2\right)
    + {2G^2S\over 15c^7\mu^2r^7}\Biggl\{
         48 Gm\mu^3Er^3+12\mu r^2\left(6 EL^2 + G^2m^2\mu^3\right)
            - 44 Gm\mu^2L^2 r                                \nonumber\\
& & \qquad\quad 4 \eta\bigl[2 \mu^2E^2r^4+12 Gm\mu^3Er^3
         + 3\mu r^2\left(5 EL^2 + G^2m^2\mu^3\right)- 5 Gm\mu^2L^2 r
         + 15 L^4\bigr]                                      \nonumber\\
&+& \sin^2\kappa \Bigl\{-24 Gm\mu^3Er^3+6\mu r^2\left(6 EL^2 -
           G^2m^2\mu^3\right) + 72 Gm\mu^2L^2 r - 90 L^4     \nonumber\\
& & \qquad\quad +\eta\bigl[-4 \mu^2E^2r^4-24 Gm\mu^3Er^3
                +6\mu r^2\left(5 EL^2 - G^2m^2\mu^3\right)
                +49 Gm\mu^2L^2 r - 135 L^4\bigr]\Bigr\}      \nonumber\\
&-& \mu Lr\dot r\sin^2\kappa\sin 2\Psi
     \Bigl\{36 \mu Er^2+12 Gm\mu^2r-18 L^2
      +\eta\left(34 \mu Er^2+12 Gm\mu^2r-15L^2\right)\Bigr\} \nonumber\\
&+& \sin^2\kappa\cos 2\Psi \Bigl\{ 24Gm\mu^3Er^3 + 6\mu r^2\left(6 EL^2
       + G^2m^2\mu^3\right) + 6 Gm\mu^2L^2 r + 18 L^4        \nonumber\\
& & \qquad\qquad\qquad +\eta\bigl[4 \mu^2E^2r^4 + 24 Gm\mu^3Er^3
          +2\mu r^2(13 EL^2+3G^2m^2\mu^3)+3 Gm\mu^2L^2 r+15 L^4
            \bigr]\Bigr\}\Biggr\}                            \nonumber\\
&+& \left({{\bf L}\over S}\cdot{d{\bf S}\over dt}\right)
   -\left({{\bf L}\over S}\cdot{{\bf S}\over S}\right)
    \left({{\bf S}\over S}\cdot{d{\bf S}\over dt}\right)
\ .\nonumber \end{eqnarray}
 
Now we proceed to evaluate the two radiative spin-loss terms
appearing in the $L_S$-loss. We need these terms to the
Newtonian order and we compute them from the radiation-reaction
potential. To lowest order, this is the Burke-Thorne potential
\cite{BT,Blanchet}
\begin{equation} \label{BTpot}
 V=-{G\over 5c^5}I_{jk}^{(5)}y_jy_k \ ,
\end{equation}
where $I_{jk}^{(5)}$ is the fifth time derivative of the system's
quadrupole-moment tensor and $y_j$ are coordinates centered on
the spinning body. The torque ${\bf \tau}$ is the
integral of the backreaction force over the volume of the body:
\begin{equation}
 \tau_i={2G\over 5c^5}\epsilon_{ikl}I_{jl}^{(5)}
                                    \int\rho(y) y_ky_j dV \ .
\end{equation}
The torque induces the radiative change in the spin. The volume
integral can be expressed with the tensor of inertia
 $\Theta_{jk}=\int \rho (y_jy_k-\delta_{jk}y_ly_l)dV$:
\begin{equation}
 {dS_i\over dt}=\tau_i
  = {2G\over 5c^5}\epsilon_{ikl}I_{jl}^{(5)}\Theta_{jk} \ .
\end{equation}
 
For an axisymmetrically spinning body the tensor of inertia is
$\Theta_{jk}=diag(\Theta,\Theta,\Theta')$ in the system of
principal axes with ${\bf S}$ on the $z$ axis.
The magnitude of the spin is $S=\Theta'\Omega$, where
$\Omega$ is the angular velocity of the body. Thus
\begin{equation} \label{Sdirdot}
 {1\over S}{dS_i\over dt} =
 {2G\over 5c^5\Omega }\bigl({\Theta\over\Theta'}-1\bigr)
   \epsilon_{ijk}I_{jl}^{(5)}{\bf\hat{S}}_k{\bf\hat{S}}_l
\end{equation}
is perpendicular to the spin ${\bf S}$. Here the quantity
$\delta=\Theta/\Theta'-1$ is the deviation from sphericity and
${\bf\hat{S}}$ the spin direction.
When the body is rapidly rotating ($V\gg \epsilon^{1/2}$), it is
centrifugally flattened ($\delta$ is nonnegligible) and the estimate
of ACST \cite{ACST} for $d{\bf S}/dt$ holds. For slow rotation,
$V\approx\epsilon^{1/2}$, the deviation from sphericity is $\delta$ is
small. In both cases, the approximation above, equating $d{\bf J}/dt$
with $d{\bf L}/dt$ is valid.
 
  From (\ref{Sdirdot}), the last term in the $L_S$-loss vanishes,
and the other term containing the radiative spin loss is
\begin{eqnarray} \label{spinlossterm}
 {{\bf L}\over S}\cdot{d{\bf S}\over dt} =
 -{2G^2mL\sin^2\kappa\over 5c^5\mu^2\Omega r^7}
             \bigl({\Theta\over\Theta'}-1\bigr)
 \Bigl\{&&\mu r\dot{r}\sin 2\Psi\left(12\mu Er^2+20Gm\mu^2r+45L^2\right)\\
   &+&4L\cos 2\Psi\left(18\mu Er^2+20Gm\mu^2r-15L^2\right)\Bigr\}
\nonumber \ .
\end{eqnarray}
 
The losses in the energy and in the magnitude of angular momentum
contain constants of the motion and they depend only on $r=r(\chi)$.
These expressions
will take the suitable form for averaging, after inserting the
parametrization (\ref{eq:chi}). The loss $dL_S/dt$ depends also on
the zeroth order expressions of the angle variable $\Psi$ and of
$\dot r$, which are:
\begin{equation}
\Psi=\Psi_0+\chi\ ,\qquad \dot r={A_0\over L}\sin\chi\ .
\end{equation}
With these expressions, the task of parametrizing is
completed: the loss in $L_S$ can also be expressed in terms of
the true anomaly parameter $\chi$.
 
\section{Averaged radiative losses}
 
The averaged losses are computed by the residue theorem, passing
to the complex variable $\zeta=e^{i\chi}$. The average of the term
(\ref{spinlossterm}) vanishes,
\begin{equation}
 \Bigl<{{\bf L}\over S}\cdot{d{\bf S}\over dt}\Bigr> = 0 \ .
\end{equation}
 
The averaged losses in the constants of motion are:
\begin{mathletters}
\begin{eqnarray} \label{averloss}
\Bigl<{dE\over dt}\Bigr>=
 &-&{G^2m(-2E\mu)^{3/2}\over 15c^5L^7}
      (148E^2L^4+732G^2m^2\mu^3EL^2 +425G^4m^4\mu^6)        \nonumber\\
 &+&{G^2(-2E\mu)^{3/2}S\cos\kappa\over 10c^7L^{10}}
      \Bigl[520E^3L^6+10740G^2m^2\mu^3E^2L^4+24990G^4m^4\mu^6EL^2
           +12579G^6m^6\mu^9               \label{averlossE}   \\
 & &\qquad\qquad\qquad\qquad\quad +\eta(256E^3L^6+6660G^2m^2\mu^3E^2L^4
      +16660G^4m^4\mu^6EL^2+8673G^6m^6\mu^9)\Bigr]          \nonumber\\
\Bigl<{dL\over dt}\Bigr>=
 &-&{4G^2m(-2E\mu)^{3/2}\over 5c^5L^4}(14EL^2+15G^2m^2\mu^3) \nonumber\\
 &+&{G^2(-2E\mu)^{3/2}S\cos\kappa\over 15c^7L^7}
      \Bigl[1188E^2L^4+6756G^2m^2\mu^3EL^2+5345G^4m^4\mu^6\label{averlossL} \\
 & &\qquad\qquad\qquad\qquad\quad
        +\eta(772E^2L^4+4476G^2m^2\mu^3EL^2+3665G^4m^4\mu^6)\Bigr]
                                                             \nonumber\\
\Bigl<{dL_S\over dt}\Bigr>=
 &-&{4G^2m(-2E\mu)^{3/2}\cos\kappa\over 5\,c^5L^4}
    (14EL^2+15G^2m^2\mu^3) + {G^2(-2E\mu)^{3/2}S\over 15\,c^7L^7} \nonumber\\
 &\Bigl\{ &\bigl[1188E^2L^4+6756G^2m^2\mu^3EL^2 + 5345G^4m^4\mu^6
      +\eta(772E^2L^4+4476G^2m^2\mu^3EL^2+3665G^4m^4\mu^6)\bigr]
                                         \nonumber \\
 &-&{\sin^2\kappa\over 2}
     \bigl[3516E^2L^4+17676G^2m^2\mu^3EL^2+12975G^4m^4\mu^6      \nonumber\\
 & &\qquad\quad
    +\eta(2428E^2L^4+12216G^2m^2\mu^3EL^2+9125G^4m^4\mu^6)\bigr] \label{averlossLS}\\
 &+&{\sin^2\kappa\cos(2\Psi_0)}
     \bigl[(26EL^2+33G^2m^2\mu^3)(12EL^2+6G^2m^2\mu^3)           \nonumber\\
 & &\qquad\qquad\qquad\quad
   +\eta
 (119EL^2+156G^2m^2\mu^3)(2EL^2+G^2m^2\mu^3)\bigr]\Bigr\}
\ .              \nonumber
\end{eqnarray}
\end{mathletters}
The expressions (\ref{averlossE}) and (\ref{averlossL}) agree with the
available results of \cite{Rieth} for the
averaged power and $<dL/dt>$, despite the differences
encountered in the higher-order terms of the definition
of the energy and angular momentum.
 
  The averaged loss in $<{dL_S/dt}>$ contains terms proportional to
$\cos(2\Psi_0)$. Ryan\cite{Ry2} argues that in most cases such
terms are averaged out by the Schwarzschild precession.

\section{Averaged Losses in Terms of Orbit Parameters}
 
As in {\bf I}, we define the generalized semimajor axis and eccentricity
imposing the conditions:
\begin{equation}\label{rmin}
r_{{}_{min}^{max}}=a(1\pm e)\ .
\end{equation}
Solving the system (\ref{turning}) and (\ref{rmin}), one has:
\begin{eqnarray} \label{a1me2}
a=&-&{Gm\mu \over 2E}\Bigl[1-{2EL_SS\over c^2mL^2}(2+\eta)\Bigr]
  \ ,\nonumber\\
1-e^2=-{2EL^2\over G^2m^2\mu^3}\Bigl[1&+&{8L_SS\over c^2mL^4}
  (EL^2+G^2m^2\mu^3)+{2\eta L_SS\over c^2mL^4}(EL^2+2G^2m^2\mu^3)\Bigr]\ .
\end{eqnarray}
From these relations one can express the constants of motion in terms of
the orbit parameters:
\begin{eqnarray} \label{el2}
E=&-&{Gm\mu \over 2\,a}\Bigl(1+{GS\cos\kappa(2+\eta)\over
                        c^2a^{3/2}\sqrt{Gm(1-e^2)}}\Bigr)\ ,\nonumber\\
L^2=Gm\mu^2 a(1&-&e^2)\Bigl(1-{2GS\cos\kappa\over
  c^2a^{3/2}\sqrt{Gm(1-e^2)}}{3+e^2+2\eta\over 1-e^2}\Bigl)\ .
\end{eqnarray}
Inserting (\ref{el2}) in the equations (\ref{averloss}) we get
the averaged radiation losses in terms of $a,\ e$ and $\kappa$:
\begin{eqnarray} \label{averlossaeka}
\Bigl<{dE\over dt}\Bigr>=
 &-&{G^4m^3\mu^2\over 15c^5a^5(1-e^2)^{7/2}}(37e^4+292e^2+96)  \nonumber\\
 &+&{G^{9/2}m^{5/2}\mu^2S\cos\kappa\over 30c^7a^{13/2}(1-e^2)^5}
 \Bigl[491e^6+5694e^4+6584e^2+1168+\eta
(355e^6+5316e^4+7248e^2+1200)\Bigr]                       \nonumber\\
\Bigl<{dL\over dt}\Bigr>=
 &-&{4G^{7/2}m^{5/2}\mu^2\over 5c^5a^{7/2}(1-e^2)^2}(7e^2+8)   \nonumber\\
 &+&{G^4m^2\mu^2S\cos\kappa\over 15c^7a^5(1-e^2)^{7/2}}
  \Bigl[549e^4+1428e^2+488+\eta (403e^4+1366e^2+456)\Bigr]            \\
\Bigl<{dL_S\over dt}\Bigr>=
 &-&{4G^{7/2}m^{5/2}\mu^2\cos\kappa\over 5c^5a^{7/2}(1-e^2)^2}(7e^2+8)
   +{G^4m^2\mu^2S\over 30c^7a^5(1-e^2)^{7/2}}                  \nonumber\\
 &\Bigl\{ &2\bigl[549e^4+1428e^2+488
        +\eta (403e^4+1366e^2+456)\bigr]        \nonumber\\
 &-&\sin^2\kappa \bigl[1383e^4+4368e^2+1464+\eta
         (1027e^4+3922e^2+1296)\bigr]                 \nonumber\\
 &+&\sin^2\kappa\cos(2\Psi_0)
     \,e^2\bigl[156e^2+240+\eta (119e^2+193)\bigr]\Bigr\}
\ .\nonumber
\end{eqnarray}
 
The losses of $\kappa, a$ and $e$ follow by taking the time
derivatives of (\ref{LS}) and of (\ref{a1me2}):
\begin{eqnarray}
\label{aekaloss} \Bigl<{d\kappa\over dt}\Bigr>=
 &+&{G^{7/2}m^{3/2}\mu S\sin\kappa\over 30c^7a^{11/2}(1-e^2)^4}
   \Bigl\{ 285e^4+1512e^2+488+\eta (221e^4+1190e^2+384)          \nonumber\\
 & &\qquad\qquad\qquad\qquad\quad
 -e^2\cos(2\Psi_0)\bigl[156e^2+240+\eta (119e^2+193)\bigr]\Bigr\}\nonumber\\
\Bigl<{da\over dt}\Bigr>=
 &-&{2G^3m^2\mu\over 15c^5a^3(1-e^2)^{7/2}}(37e^4+292e^2+96)              \\
 &+&{G^{7/2}m^{3/2}\mu S\cos\kappa\over 15c^7a^{9/2}(1-e^2)^5}
  \Bigl[363e^6+3510e^4+7936e^2+2128+\eta
   (291e^6+4224e^4+7924e^2+1680)\Bigr]               \nonumber\\
\Bigl<{de\over dt}\Bigr>=
 &-&{G^3m^2\mu\over 15c^5a^4(1-e^2)^{5/2}}\, e (121e^2+304)      \nonumber\\
 &+&{G^{7/2}m^{3/2}\mu S\cos\kappa\over 30c^7a^{11/2}(1-e^2)^4}\, e
  \Bigl[1313e^4+5592e^2+7032+\eta (1097e^4+6822e^2+6200)\Bigr]\
. \nonumber
\end{eqnarray}
From Eq. (\ref{aekaloss}), the angle $\kappa$ increases. Hence
we find that the vectors $\bf L$ and $\bf S$ of a finite-mass binary
system tend to antialign as in {\bf I}.
Substituting $\eta=0$ for a test particle with negligible mass,
we reproduce the Lense-Thirring losses of {\bf I}.
 
   The evolution of the angles $\iota$ and $\iota_N$ under radiation
backreaction is determined by $<dL/dt>$. This computation, relegated to
Appendix B, involves
higher-order terms in $S$ which are, however of (post)$^{3/2}$-order.
 
\section{Concluding Remarks}
  The computation of finite-mass effects has proved to be much more
difficult than the corresponding description of a test particle. We have
been helped by several fortunate circumstances in overcoming the
obstacles in computing the radiative losses up to the
leading spin terms. First, in this approximation, one can still
decouple a radial equation of motion.
Furthermore, a generalization of the true anomaly parametrization
$\chi$, found in {\bf I} for the Lense-Thirring case, exists with the
inclusion of the finite mass contributions. Although the dynamics
of the system is much more complicated, we succeeded in describing all
quantities entering in the radiative losses in terms of the parameter
$\chi$.
 
We were able to separately describe the relative evolution of the spin
and orbital angular momentum vectors, up to first order in spin, under
radiation backreaction. In
the loss of the spin projection of the orbital angular momentum two
terms, containing the radiative spin loss, appear. One of them is
known to vanish (ACST). We computed the other term, and have shown that
it averages out to nothing. We did not consider no-spin post-Newtonian
effects
here. In the equations of motion, we did not keep terms of
higher order than $\epsilon^{3/2}$, which, however, are comparable or
even larger than the radiative losses. They are needed for a complete
characterization of the orbit, but they do not contribute to the
radiative losses in the order to which we are computing them.
 
In the test-particle ($\eta=0$) and constant-spin case our results
agree with those of {\bf I}.
The Lense-Thirring description implies $S\gg L$.
These pictures arise as different limits of a more general
description in which both $S$ and $L$ are arbitrary.
Such a description is subject to further investigation as it will
go beyond considering only the linear terms in $S$.

\section{Acknowledgments}
 
  This work has been supported by OTKA no. T17176 and D23744
grants. The algebraic package REDUCE was used for checking our
computations. We thank the referee for amendments on the
manuscript.

\appendix
\section{Nonradiative evolution of $L_S$ and $\kappa$}
   In this Appendix, we consider the nonradiative evolution of $L_S$
and the angle $\kappa$ to linear order in the spin, and we show
that these changes are smaller than
order-$\epsilon^{3/2}$.
 
 From the conservation of the total angular momentum ${\bf J}$ and the
spin precession equation (\ref{eq:Sprec}), the time derivative of
$L_S={\bf L}\cdot{{\bf\hat S}}$ has the form:
\begin{equation}
\dot L_S=
    \frac{4+3\eta}{2r^3}\frac{G^2\mu^2S}{c^4}\Biggl\{{{\bf r\cdot v}}
    \Bigl[\frac{2+\eta}{r^3}\Bigl({{\bf r\cdot}}\frac{{\bf S}}{S}\Bigr)^2
    +\frac{\eta}{2Gm}\Bigl({{\bf v\cdot}}\frac{{\bf S}}{S}\Bigr)^2\Bigr]
    -\Bigl({{\bf r\cdot}}\frac{{\bf S}}{S}\Bigr)
     \Bigl({{\bf v\cdot}}\frac{{\bf S}}{S}\Bigr)
     \Bigl(\frac{2+\eta}{r}+\frac{\eta v^2}{2Gm}\Bigr)\Biggr\}\ .
\end{equation}
The multiplicative factor of $S$ needs to be evaluated only to Newtonian
order. We write the non-radiative change in $L_S$ in terms of the
radial and Euler-angle coordinates (\ref{eq:xyz}):
\begin{equation}
\dot L_S=
   -{GS(4+3\eta)\sin^2\kappa\over 4c^4m\mu r^5}\Bigl[
     L\bigl(-E\eta\mu r^2+2Gm\mu^2 r+L^2\eta\bigr)\sin(2\Psi)
          -L^2\dot{r}\eta\mu r\cos(2\Psi)   \Bigr] \ .
\end{equation}
After parameterizing by $\chi$, the averaged expression over one period
is:
\begin{equation}
\label{eq:Lsdot}
\bigl<\dot L_S\bigr>=
   -{G(-2E\mu)^{3/2}S\mu A_0^2\sin^2\kappa\sin(2\Psi_0)\over 16c^4L^4m}
           (15\eta^2+26\eta+8) \ .
\end{equation}
  This expression is of order $r\epsilon^{9/2}V$, comparable with the
leading radiative loss in $L$ given by Peters and Mathews, corresponding
to 2.5PN relative order.
However we pursue our description to the 1.5PN relative order.
If further terms were considered, e.g. $\epsilon^{5/2}$, they would
generate additional terms in ${\bf L}$, as well as in $\dot L_S$.
It is simple to check that these terms would be of the same order as
(\ref{eq:Lsdot}). Thus the change in $L_S$ is
negligible in our approximation.
 
  Similarly, the change in $\cos\kappa$ is negligible, as follows from
$(\cos\kappa){\bf\dot{}}={\dot L_S/ L}$.
 
\section{Evolution of $\iota$ and $\iota_N$}
 In this Appendix, we discuss the evolution,
both in the absence of radiation and under radiation
backreaction, of the angles
\begin{equation}\label{eq:defio}
\cos\iota=\frac{{\bf L}}{L}\cdot\frac{{\bf J}}{J}\qquad
\cos\iota_N=\frac{{\bf L_N}}{L_N}\cdot\frac{{\bf J}}{J}\ ,
\end{equation}
where $J$ and $L_N$ are the magnitude of the total angular momentum
${\bf J}$ and ${\bf L_N}$, respectively.
 
   An expansion of the expressions on the right hand sides of
(\ref{eq:defio}) to 1.5PN order yields
\begin{equation}
\cos\iota=\cos\iota_N=1-\frac{1}{2}\sin^2\kappa\left(\frac{S}{L}\right)^2
                     -\cos\kappa\sin^2\kappa\left(\frac{S}{L}\right)^3\ .
\end{equation}
  By this relation,
both the radiative and the nonradiative changes of $\cos\iota$ and
$\cos\iota_N$ are given in terms of the corresponding changes in
$\kappa$, $S$ and $L$:
\begin{eqnarray}  \label{eq:deioion}
\delta\cos\iota=\delta\cos\iota_N
    &=&-\left(\frac{S}{L}\right)^2
             \biggl[\cos\kappa\,\delta\cos\kappa+\sin^2\kappa
               \biggl(\frac{\delta S}{S}-\frac{\delta L}{L}\biggr)\biggr]
                                               \nonumber\\
    &&-\left(\frac{S}{L}\right)^3
         \biggl[\bigl(1-3\cos^2\kappa\bigr)\delta\cos\kappa
          +3\cos\kappa\sin^2\kappa
           \biggl(\frac{\delta S}{S}-\frac{\delta L}{L}\biggr)\biggr]\ .
\end{eqnarray}
The first and second terms on the right hand side are of 1PN order
and of 1.5PN order, respectively.
 
We have shown in Appendix A that the nonradiative evolution of the
angle $\kappa$ is negligible in our 1.5PN approximation. Because $L$ and
$S$ are also constants, $(\cos\iota)\dot{}=\cos\iota_N\dot{}=0$
under the nonradiative evolution.
 
Finally we turn our attention to the radiative evolution of the angles
$\iota$ and $\iota_N$.
The spin term correction in (\ref{averlossaeka}) to the leading order
radiative loss of $L$ and the loss (\ref{aekaloss}) of the angle
$\kappa$ are of 1.5PN order, as they originate in the spin-orbit term
$L_{SO}$. Thus these terms do not contribute to the losses
(\ref{eq:deioion}). Similarly the loss in the magnitude of the spin $S$
was seen to vanish by (\ref{Sdirdot}). Hence the only loss
contributing to the radiative changes in $\iota$ and $\iota_N$ is
the leading term in the loss of $L$:
\begin{equation}
\frac{d \cos\iota}{dt}=\frac{d\cos\iota_N}{dt}
     =\left(\frac{S}{L}\right)^2
             \biggl(1+3\cos\kappa\frac{S}{L}\biggr)
      \frac{1}{L}\frac{d L}{dt}\ .
\end{equation}
  As the factors multiplying $dL/dt$ are constant during a period of
revolution, the averaged rates can be simply obtained by use of the
leading terms in Eq. (\ref{averlossL}) for $<dL/dt>$.


\begin{thebibliography}{99}
\bibitem{c1}We adopt the convention that the derivatives with respect
to the time parameter $t$ occurring in the equations of motion
(Sec.II) are denoted by an overdot but we reserve
the notation $d/dt$ for time derivatives due to radiation backreaction.
\bibitem{c2}The gravitational
backreaction will slowly rotate out the vector $J$ from the $z$
direction. In our perturbative treatment, this effect, as well as the
changes in the other constants of motion are small during the
period of averaging, thus neglected.
\end{thebibliography}
\end{document}